\documentclass[apjl]{emulateapj}

\usepackage{graphicx}
\usepackage{amsmath}
\usepackage{natbib}

\usepackage{color}

\begin{document}

\title{
  A hot Jupiter for breakfast? --- Early stellar ingestion of planets may be
  common
}

\author{Titos Matsakos and Arieh K\"onigl}

\affil{
  Department of Astronomy \& Astrophysics and The Enrico Fermi Institute,
  The University of Chicago, Chicago, IL 60637, USA
}

\shortauthors{Matsakos \& K\"onigl}
\shorttitle{Early Stellar Ingestion of Planets}

\begin{abstract}
Models of planet formation and evolution predict that giant planets form
efficiently in protoplanetary disks, that most of these migrate rapidly to the
disk's inner edge, and that, if the arriving planet's mass is
$\lesssim$~Jupiter's mass, it could remain stranded near that radius.
We argue that such planets would be ingested by tidal interaction with the host
star on a timescale $\lesssim1\,$Gyr, and that, in the case of a solar-type
host, this would cause the stellar spin to approach the direction of the
ingested planet's orbital axis even if the two were initially highly misaligned.
Primordially misaligned stars whose effective temperatures are $\gtrsim6250\,$K
cannot be realigned in this way because, in contrast with solar-type hosts,
their angular momenta are typically higher than the orbital angular momentum of
the ingested planet as a result of inefficient magnetic braking and of a
comparatively large moment of inertia.
Hot Jupiters located farther out from the star can contribute to this process,
but their effect is weaker because the tidal interaction strength decreases
rapidly with increasing semimajor axis.
We demonstrate that, if $\sim50\%$ of planetary systems harbored a stranded hot
Jupiter, this scenario can in principle account for (1) the good alignment
exhibited by planets around cool stars irrespective of the planet's mass or
orbital period, (2) the prevalence of misaligned planets around hot stars, (3)
the apparent upper bound on the mass of hot Jupiters on retrograde orbits, and
(4) the inverse correlation between stellar spin periods and hot-Jupiter masses.
\end{abstract} 


\maketitle

\section{Introduction}
\label{sec:intro}

One of the puzzling findings in the study of exoplanets has been the difference
in the properties of planets orbiting cool and hot stars.
In particular, using measurements of the Rossiter--McLaughlin effect, it has
been inferred that the apparent obliquity $\lambda$ (the angle between the
stellar spin and planet's orbital angular momentum vectors as projected on the
sky) is typically small ($\lambda<20^\circ$) for cool stars, but that stars of
effective temperature $T_\mathrm{eff}\gtrsim6250\,$K exhibit a broad range of
apparent obliquities, reaching all the way up to $\lambda\sim180^\circ$
\citep[e.g.,][]{Winn+10,Albrecht+12}.
It was further found that the masses $M_\mathrm{p}$ of planets on apparent
retrograde orbits ($\lambda>90^\circ$) are $<3\,M_\mathrm{J}$ (where
$M_\mathrm{J}$ is Jupiter's mass; \citealt{Hebrard+11}), and that the stellar
rotation periods $P_*$ of cool stars that host a hot Jupiter (HJ) decrease with
increasing $M_\mathrm{p}$ \citep{Dawson14}.

Several models have been advanced to explain the obliquity dichotomy, but so far
none can account for all the relevant observations.
Interpretations based solely on differences in the stellar properties
\citep[e.g.,][]{Rogers+12} cannot address the inferred dependence of the
obliquity on $M_\mathrm{p}$ and the detection of highly misaligned planets
around cool stars.
In view of the fact that $T_\mathrm{eff}\approx6250\;$K corresponds to the
temperature above which the size of the outer convective zone of a main-sequence
(MS) F~star shrinks rapidly, it was suggested that close-in planets in both cool
and hot stars are initially distributed over the entire angular range
($0-180\;$degrees), but that only in cool stars where a substantial convective
envelope is present can a sufficiently massive close-in planet
($M_\mathrm{p}\gtrsim1\,M_\mathrm{J}$) realign the star through tidal
interaction \citep[e.g.,][]{Winn+10,Albrecht+12}.
The host stars are also subject to magnetic braking, which declines strongly
above the same transition temperature (corresponding to the break in the Kraft
curve; \citealt {Kraft67}), and it was argued \citep{Dawson14} that this, rather
than the tidal dissipation efficiency, is the main factor underlying the
difference in obliquity properties between cool and hot stars.
Both variants of this scenario, however, face the conundrum that an HJ
undergoing equilibrium tidal interaction with its host star would spiral in and
be ingested on a timescale that is comparable to the realignment time.
A possible way out of this difficulty is to appeal to the \{10\} component of
the dynamical tide, which could in principle significantly reduce the alignment
time without affecting the ingestion time \citep{Lai12}.\footnote{
An alternative possibility, that only the outer convective layer \citep{Winn+10}
--- or even just a part of it \citep{Dawson14} --- partakes in the realignment
process, is hard to justify on either theoretical or observational grounds.}
However, even though this model can be used to account for individual systems
\citep{ValsecchiRasio14}, it remains unclear whether it can explain the overall
$\lambda$ distribution of HJs and the manifested difference between cool and hot
stars \citep{RogersLin13,Xue+14}.
Furthermore, even the basic tidal interaction interpretation of the obliquity
dichotomy has now been called into question by the results of \citet{Mazeh+15},
who analyzed the rotational photometric modulations of a large sample of
\textit{Kepler} sources and inferred that (1) the conclusion that planets around
cool stars are well aligned, and those around hot stars are not, is general and
not restricted just to HJs; and (2) the low obliquity of planets around cool
stars extends to orbital periods $P_\mathrm{orb}$ that are a factor of $\sim10$
larger than the maximum value ($<4\;$days) for robust tidal interaction between
an HJ and a $\gtrsim1\,$Gyr-old G or F~star. 

In this Letter we propose to address these apparent difficulties and account for
many of the observed differences between the properties of planets in cool and
hot stars by postulating that, in addition to the tidal interaction with
existing close-in planets, a large fraction of the stellar hosts --- both cool
and hot --- ingest a hot Jupiter early on in their evolution.
This proposal is motivated by the expectation that a large fraction (up to
$80\%$ according to \citealt{Trilling+02}) of solar-type stars possess giant
planets during their pre-MS phase, and that a large fraction of the giant
planets that form in a protoplanetary disk on scales $\lesssim5\;$AU migrate
close to their host star before the disk is dispersed \citep[e.g.,][]{IdaLin04}.
Numerical simulations incorporating an N-body code and a 1D $\alpha$-viscosity
disk model \citep{Thommes+08} demonstrated that this behavior can be expected
for disks with $\alpha\lesssim0.01$ that are sufficiently massive.
The inward planet migration is likely stopped by the strong ($\gtrsim1\;$kG)
protostellar magnetic field that truncates the disk at a radius
($r_\mathrm{in}$) of a few stellar radii \citep[e.g.,][]{Lin+96}.
Gravitational interaction with the disk causes a planet reaching $r_\mathrm{in}$
to penetrate into the magnetospheric cavity and, if it is massive enough, to
undergo eccentricity excitation that can rapidly lead to a collision with the
star \citep[e.g.,][]{Rice+08}.
It was, however, inferred that if $M_\mathrm{p}$ is sufficiently small
($\lesssim1\,M_\mathrm{J}$), the planet would remain stranded at a distance
where its orbital period is $\sim0.5$ of that at $r_\mathrm{in}$ until well
after the gas disk disappears (on a timescale of $\sim10^6-10^7\;$yr).
In our proposed scenario, the primordial disk orientations span a broad angular
range that is reflected in the orbital orientations of the stranded planets.
When the latter are ingested by tidal interaction with the host star (on a
timescale $<1\,$Gyr), the absorbed angular momentum is sufficient to align a
solar-mass star in that general direction, but not an MS star with
$T_\mathrm{eff}\gtrsim6250\;$K.
This is because cool stars have significantly lower angular momenta at the time
of ingestion than hot stars as a result of a more efficient magnetic braking
process and of a lower moment of inertia.
Given the proximity of the stranded HJs (SHJs) to their host stars
($P_\mathrm{orb,SHJ}\lesssim2\,$days), they can be expected to have been
ingested by the time their parent planetary systems are observed; however, giant
planets farther out can continue to interact with their host stars and
potentially affect their measured obliquities.
In our simplified formulation, we model the SHJs using as parameters their
characteristic mass $M_\mathrm{SHJ}$  and the fraction $p$ of systems that
initially harbored an SHJ.
By comparing the predictions of this model with the observational data, we infer
$M_\mathrm{SHJ}\lesssim1\,M_\mathrm{J}$ and $p\sim0.5$.

The potential effect of tidally induced ingestion of HJs on the observed
properties of planetary systems has been recognized before.
In particular, \citet{Jackson+09} suggested that this process could account for
the observed orbital distribution of close-in planets, whereas
\citet{TeitlerKonigl14} proposed that the spinup induced by the deposition of a
swallowed planet's orbital angular momentum in the host's envelope can explain
the observed dearth of close-in planets around fast-rotating stars.
The current proposal further extends this scenario by incorporating not only the
distribution of observed planets but also a putative population of HJs that were
ingested after being stranded near the inner edges of their associated
protoplanetary disks.

\section{Modeling Approach}

Using Monte Carlo simulations, we follow the temporal evolution of the stellar
and orbital angular momenta of $10^6$ systems in the context of an
equilibrium-tide model.
The Cartesian coordinate frame ($x$, $y$, $z$) is chosen so that the orbital
angular momentum ($\boldsymbol{L}$) always points along the $z$ axis and the
stellar one ($\boldsymbol{S}$) always lies in the $y$-$z$ plane.
We assume circular orbits and neglect the precession of $\boldsymbol{L}$ and
$\boldsymbol{S}$ around the total angular momentum vector.\footnote{
The associated torques act along the $x$ axis and thus do not affect either the
orbital separation or the alignment.}
The set of equations employed to model the tidal interaction between a star
(subscript~$*$) and a planet (subscript p) is:
\begin{equation}
  \frac{d\boldsymbol L}{dt}=+\frac{L}{2\tau_\mathrm{d}}\frac{\Omega_{*x}}{\Omega_\mathrm{orb}}\hat{\boldsymbol x}+\frac{L}{2\tau_\mathrm{d}}\frac{\Omega_{*y}}{\Omega_\mathrm{orb}}\hat{\boldsymbol y}-\frac{L}{\tau_\mathrm{d}}\left(1-\frac{\Omega_{*z}}{\Omega_\mathrm{orb}}\right)\hat{\boldsymbol z}\,,
\end{equation}
\begin{equation}
  \frac{d\boldsymbol S}{dt}=-\frac{L}{2\tau_\mathrm{d}}\frac{\Omega_{*x}}{\Omega_\mathrm{orb}}\hat{\boldsymbol x}-\frac{L}{2\tau_\mathrm{d}}\frac{\Omega_{*y}}{\Omega_\mathrm{orb}}\hat{\boldsymbol y}+\frac{L}{\tau_\mathrm{d}}\left(1-\frac{\Omega_{*z}}{\Omega_\mathrm{orb}}\right)\hat{\boldsymbol z}-\boldsymbol T\,,
\end{equation}
where $\boldsymbol\Omega_\mathrm{orb}=\Omega_\mathrm{orb}\,\hat{\boldsymbol{z}}$
and $\boldsymbol\Omega_*$ are the orbital and stellar angular velocities,
respectively (related to the orbital and stellar rotation periods through
$P_\mathrm{orb}=2\pi/\Omega_\mathrm{orb}$ and $P_*=2\pi/\Omega_*$),
$\boldsymbol{T}=T\hat{\boldsymbol S}$ is the magnetic braking torque, and
$\tau_\mathrm{d}$ is the nominal tidal damping time, which, neglecting
dissipation inside the planet, is given by
\begin{equation}
  \tau_\mathrm{d}=\frac{4Q^\prime_*}{9}\left(\frac{a}{R_*}\right)^5\frac{M_*}{M_\mathrm{p}}\frac{1}{\Omega_\mathrm{orb}}\,.
\label{eq:tau}
\end{equation}
In Equation~(\ref{eq:tau}), $Q^\prime_*$ is the tidal quality factor (taken to
be a constant, for simplicity), $a$ is the semimajor axis, whereas $R$ and $M$
denote an object's radius and mass, respectively.
We model the star as a uniformly rotating body ($S = I_* \Omega_*$) with a
moment of inertia $I_*\approx0.06\,M_*R_*^2$, and adopt for $T$ the expression
presented by \cite{Matt+15}, which covers both the saturated and the unsaturated
regimes and incorporates as a key variable the convective turnover timescale as
a function of $T_\mathrm{eff}$ ($3300\leq T_\mathrm{eff}\leq7000$;
\citealt{CranmerSaar11}).
The nominal magnetic braking time is $\tau_\mathrm{mb} = |S/T|$.

The equations are solved in a two-step process: we first integrate for a small
$\Delta t$ employing a 4th-order Runge-Kutta method and then rotate the vectors
so that the updated $\boldsymbol{L}$ is along $\hat{\boldsymbol{z}}$ and the
updated $\boldsymbol{S}$ lies in the $y$-$z$ plane.
When considering a multiplanet system, we solve the equations for the different
planets in sequence at each time step, neglecting possible interactions between
them and assuming that their orbits remain coplanar.

\begin{deluxetable}{lcc}
\tablecaption{Model Parameters\label{tab:parameters}}
\tablehead{\colhead{Parameter}                        & \colhead{Value (G / F)}                   & \colhead{Distribution}}
\startdata
$T_\mathrm{eff}\,[\mathrm{K}]$                        & $5500$ / $6400$                           &                             \\
$M_*\,[M_{\sun}]$                                     & $1.0$ / $1.3$                             &                             \\
$R_*\,[R_{\sun}]$                                     & $1.0$ / $1.3$                             &                             \\
$P_{*\mathrm{ini}}\,[\mathrm{days}]$                  & $5$--$10$ / $2.5$--$5$\tablenotemark{a}   & Uniform                     \\
Age [Gyr]                                             & $1$--$8$ / $1$--$4$\tablenotemark{b}      & Fit to data                 \\
\hline
Number of planets                                     & $5$                                       &                             \\
$P_{\mathrm{orb}}\,[\mathrm{days}]$                   & $0.5$--$50$\tablenotemark{c}              & $f(\ln P) \propto P^{0.54}$ \\
$R_\mathrm{p}\,[R_\oplus]$ for $P_\mathrm{orb} < 7\,\mathrm{d}$
                                                      & $2$--$20$\tablenotemark{c}                &
                                                                             $f(\ln R_\mathrm{p}) \propto R_\mathrm{p}^{-1.09}$ \\
$R_\mathrm{p}\,[R_\oplus]$ for $P_\mathrm{orb} > 7\,\mathrm{d}$
                                                      & $2$--$20$\tablenotemark{c}                &
                                                                             $f(\ln R_\mathrm{p}) \propto R_\mathrm{p}^{-2.31}$ \\
$M_\mathrm{p}\,[M_\oplus]$ for $R_\mathrm{p} < 9R_\oplus$
                                                      & $4$--$81$\tablenotemark{d}                &
                                                                                            $(R_\mathrm{p}/R_\oplus)^2M_\oplus$ \\
$M_\mathrm{p}\,[M_\oplus]$ for $R_\mathrm{p} > 9R_\oplus$
                                                      & $56$--$5620$\tablenotemark{d}             & Fit to data                 \\
$\psi_\mathrm{ini}\,[\mathrm{degrees}]$ (random)      & $0$--$180$                                & $f(\psi)\propto\sin(\psi)$  \\
\hline
$p$ (SHJ fraction)                                    & $0.5$                                     &                             \\
$M_\mathrm{SHJ}\,[M_\mathrm{J}]$                      & $0.6$                                     &                             \\
$R_\mathrm{SHJ}\,[R_\mathrm{J}]$                      & $1.0$                                     &                             \\
$P_{\mathrm{orb,SHJ}}\,[\mathrm{days}]$               & $2.0$                                     &
\enddata
\tablecomments{The subscript ``ini'' indicates an initial value.}
\tablerefs{
  $^\mathrm{a}$~\cite{Meibom+11};
  $^\mathrm{b}$~\cite{WalkowiczBasri13};
  $^\mathrm{c}$~\cite{Youdin11};
  $^\mathrm{d}$~\cite{Weiss+13}.
}
\end{deluxetable}

We consider two sets of models based on the type of host star: G or F; see
Table~\ref{tab:parameters} for a listing of the adopted parameters.
For each stellar type, we evolve $5\times10^5$ systems, in each case considering
simultaneously the SHJs (parametrized by $M_\mathrm{SHJ}$ and $p$; see
Section~\ref{sec:intro}) as well as $5$~additional planets,\footnote{
Our results are not sensitive to the choice of the number of additional planets
because large, close-in planets --- which, according to Equation~(\ref{eq:tau}),
exert the strongest tidal effect --- are a rare occurrence.}
for which we randomly pick the orbital periods, planet radii, and system age
according to the procedure described in \cite{TeitlerKonigl14}.\footnote{
In relating planet masses to planet radii, we generalized the treatment in
\cite{TeitlerKonigl14} by employing a power-law relationship only for
$R_\mathrm{p}<9\,R_\earth$ and using an empirical fit to the data for
$R_\mathrm{p}>9R_\earth$; see Table~\ref{tab:parameters}.}
As in that work, we adopt the approximation that the entire planetary orbital
angular momentum is added to $\boldsymbol{S}$ when the planet reaches the Roche
limit $a_\mathrm{R}\approx1.5\,(3\,M_*/M_\mathrm{p})^{1/3}R_\mathrm{p}$, and we
neglect stellar evolution effects (but see, e.g., \citealt{Valsecchi+15} and
\citealt{ValsecchiRasio14}).
The effective value of $Q^\prime_*$ and its physical basis are still open
questions \citep[e.g.,][]{Ogilvie14}.
Note, however, that the apparent break at $P_\mathrm{orb}\gtrsim3\,$days in the
period distribution of planets around solar-mass stars
\citep[e.g.,][]{Youdin11}, if due to tidal interaction, is consistent with
$Q^\prime_*\approx10^6$ \citep[e.g.,][]{TeitlerKonigl14}.
Using the data from \texttt{exoplanets.org}, we find that F-star systems
manifest a similar break at the same location.
We therefore adopt $Q^\prime_*\sim10^6$ for both our cool and hot fiducial
stars.

The initial spin--orbit angle $\psi_\textrm{ini}$ is taken to be random,
$f(\psi_\mathrm{ini})\propto\sin(\psi_\mathrm{ini})$, which corresponds to a
flat distribution for the projected angle $\lambda_\mathrm{ini}$
\citep[see][]{FabryckyWinn09}.
Our choice of $P_\mathrm{orb,SHJ}\approx2\,$days is motivated by the
characteristic period ($\sim7\,$days; \citealt{GalletBouvier13}) of
disk-accreting protostars, which can be interpreted in terms of the magnetic
disk-locking model \citep[e.g.,][]{Konigl91}.
In this picture, the disk truncation radius is located at
$r_\mathrm{in}\approx0.7\,r_\mathrm{co}$ (with $r_\mathrm{co}$ defined by
$\Omega_\mathrm{orb}(r_\mathrm{co})=\Omega_*$) if the magnetic interaction is
dominated by the dipolar field component \citep{Long+05}.
The choice of $P_\mathrm{orb,SHJ}$ then follows from the results of
\cite{Rice+08}, who inferred that it is
$\sim0.5\,P_\mathrm{orb}(r_\mathrm{in})$.\footnote{
Using the numerical results of \cite{Long+05} and the characteristic radius
($\sim2\,R_\sun$) and mass accretion rate
($\sim10^{-8}\,M_\sun\,\mathrm{yr}^{-1}$) for classical T Tauri stars, we infer
a stellar dipolar field strength of $\sim2.5\,$kG, which is consistent with
observations \citep[e.g.,][]{Johns-Krull07}.}

\section{Results}

\begin{figure*}
  \includegraphics[width=\textwidth]{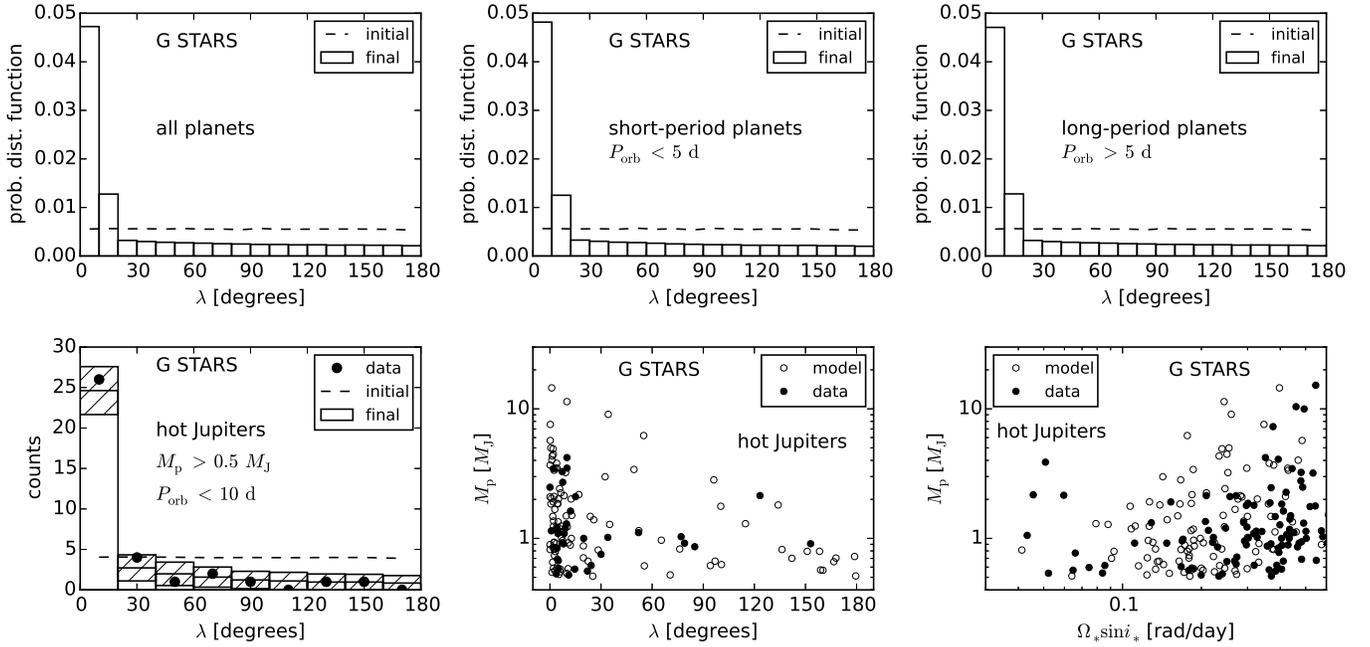}
  \caption{
    System properties in the cool-star model, as functions of either the
    projected obliquity $\lambda$ or the projected stellar angular velocity.
    The top panels include the entire range of planet masses and show the
    $\lambda$ probability distribution for all the planets (top left) and
    separately for those with $P_\mathrm{orb}<5\,\mathrm{days}$ (top middle) and
    $P_\mathrm{orb}>5\,\mathrm{days}$ (top right).
    The bottom panels show results for hot Jupiters only, along with
    observational data taken from \texttt{exoplanets.org}: the lower left
    displays planet number counts and the other two the planet mass.
    \label{fig:cool_stars}}
\end{figure*}

\begin{figure*}
  \includegraphics[width=\textwidth]{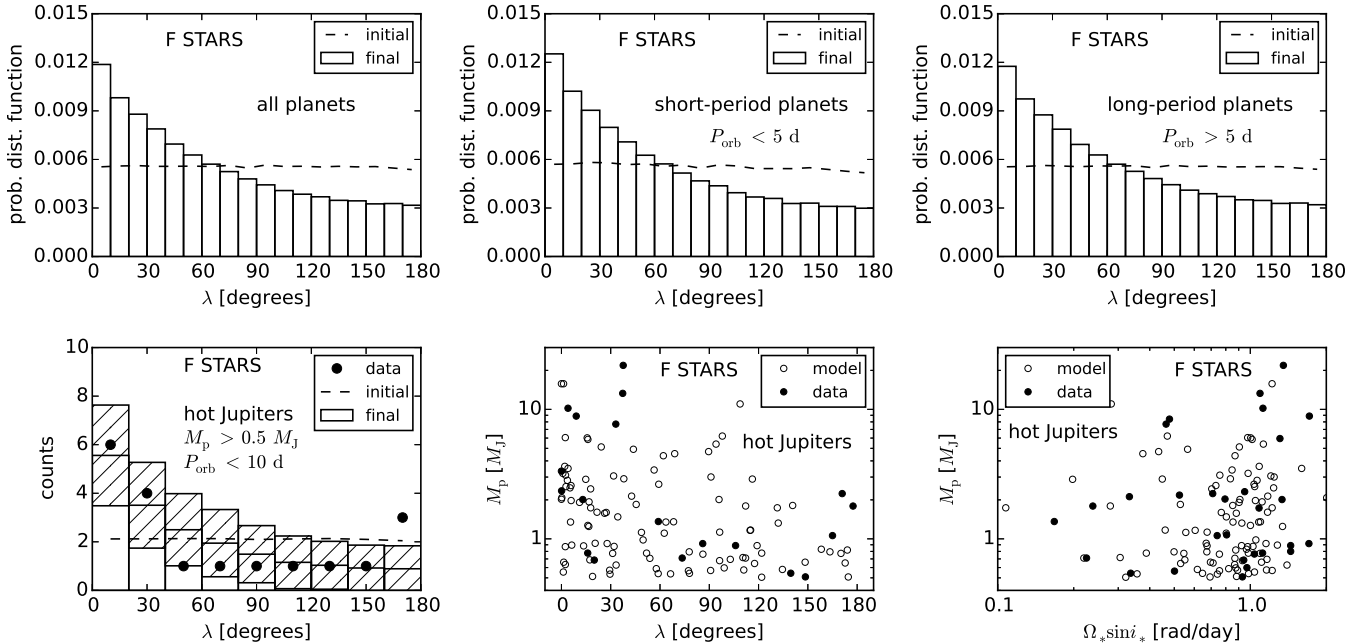}
  \caption{Same as Fig.~\ref{fig:cool_stars}, but for the hot-star model.
    \label{fig:hot_stars}}
\end{figure*}

The results for the final configurations of the modeled planetary systems are
presented in identical formats in Figures~\ref{fig:cool_stars}
and~\ref{fig:hot_stars} for our representative G and F stars, respectively.
The top three panels show the probability distribution function of the projected
spin--orbit angle for the entire population (left) and separately for its
short-period ($P_\mathrm{orb}<5\,\mathrm{days}$, middle) and long-period
($P_\mathrm{orb}>5\,\mathrm{days}$, right) components.
These panels demonstrate that our model can account for the basic trends
uncovered by the \cite{Mazeh+15} study --- that planets around cool stars are
well aligned with their host's spin irrespective of the planet's size or orbital
period, but that planets around hot stars are not.
To quantify the degree of alignment, we follow \cite{Mazeh+15} and evaluate the
ratio $q\equiv\left<\sin(i_*)\right>/(\pi/4)$, where $i_*\in[0,\,\pi/2]$ is the
angle between the stellar rotation axis and the line of sight, the angle
brackets denote an average over the distribution, and the normalization is by
the value of $\left<\sin(i_*)\right>$ for a random orientation of
$\boldsymbol{S}$ on the plane of the sky.
This ratio is therefore $1$~when there is no preferred spin--orbit orientation
(as in our adopted initial condition), whereas it is $q=1/(\pi/4)\simeq1.273$
when $\boldsymbol{S}$ is normal to the line of sight (which, in the case of a
transiting planet, likely corresponds to $\lambda\approx0$).
For our G-host model we obtain $q=1.145$, which corresponds to having the
$\sim50\%$ of systems that ingested an SHJ attain good alignment even as the
rest remain close to their initial random orientations.
For the F-host systems we find $q=1.030$, indicating that in this case the
initial random distribution is only slightly modified.
We note from Figure~1 of \cite{Mazeh+15} that, while the median value of $q$ for
stars with $T_\mathrm{eff}\lesssim5000\,$K is consistent with perfect alignment,
this ratio decreases toward $1$ in the G-star range
($T_\mathrm{eff}\sim5300-6000\,$K) that we have modeled.\footnote{
\cite{Mazeh+15} inferred $q<1$ for their hot-star sample, a value that cannot be
explained in the context of our basic model.
Note in this connection that, as discussed by these authors, the observational
challenges that need to be overcome in deriving the hot-star result are more
severe than those encountered in the cool-star case.}

The bottom three panels of Figures~\ref{fig:cool_stars} and~\ref{fig:hot_stars}
confront our model predictions for HJs (defined here by
$M_\mathrm{p}>0.5\,M_\mathrm{J}$ and $P_\mathrm{orb}<10\,\mathrm{days}$) with
the observational data (as of June 2015) obtained from \texttt{exoplanets.org},
which consist of $N_\mathrm{G}=36$ entries for cool stars
($T_\mathrm{eff}<6250\,\mathrm{K}$) and $N_\mathrm{F}=19$ entries for hot ones
($6250\,\mathrm{K}<T_\mathrm{eff}<7000\,\mathrm{K}$).
The left panel shows the planet counts as a function of $\lambda$, with the
model predictions obtained by selecting $1000$ samples from our simulations,
each consisting of $N_\mathrm{G}$ (Figure~\ref{fig:cool_stars}) or
$N_\mathrm{F}$ (Figure~\ref{fig:hot_stars}) systems, and plotting the average
counts per bin (bars) as well as the range of $\pm1$ standard deviation (shaded
areas).
It is seen that the model is in excellent agreement with the data, with the only
$>1\sigma$ deviation exhibited by the bin encompassing
$\lambda=180^\circ$.\footnote{
One possible explanation of the observed excess --- if indeed it is significant
--- is that it arises from the action of the \{10\} component of the dynamical
tide induced by surviving HJs, which has been neglected in our treatment
\citep[see][]{Xue+14}.}

\begin{table}
\caption{SHJ Ingestion\label{tab:ingestion}}
\begin{tabular}{l|ccc|ccc}
\hline\hline
Parameter                          &        & G star &         &        & F star &         \\
                                   & Init.  & Roche  & Ingest. & Init.  & Roche  & Ingest. \\
\hline
$t\,\mathrm{[Gyr]}$                & $0.0$  & $0.69$ & $0.69$  & $0.0$  & $0.35$ & $0.35$  \\
$a\,\mathrm{[AU]}$                 & $0.03$ & $0.01$ & ---     & $0.03$ & $0.01$ & ---     \\
$P_\mathrm{orb}\,\mathrm{[days]}$  & $2$    & $0.5$  & ---     & $2$    & $0.5$  & ---     \\
$P_*\,\mathrm{[days]}$             & $7.5$  & $12.9$ & $5.0$   & $3.8$  & $4.5$  & $4.2$   \\
$\tau_\mathrm{mb}\,\mathrm{[Gyr]}$ & $0.21$ & $0.61$ & $0.09$  & $1.96$ & $2.83$ & $2.46$  \\
$|L_\mathrm{SHJ}/S|$               & $1.59$ & $1.71$ & ---     & $0.43$ & $0.33$ & ---     \\
$\psi\,\mathrm{[degrees]}$         & $100$  & $35$   & $13$    & $100$  & $86$   & $69$    \\
\hline
\end{tabular}
\end{table}

The key feature that enables our model to account for the aforementioned
observations is the ingestion, over time intervals ($t_\mathrm{ingest}$) that
are shorter than the ages of the observed systems, of SHJs with values of
$|L_\mathrm{SHJ}/S|$ that are $>1$ for G stars and $<1$ for F stars (see
Table~\ref{tab:ingestion}).
The low obliquity attained by initially misaligned G stars at
$t_\mathrm{ingest}$ is due mainly to the efficient reduction of $S$ through
magnetic braking over the time interval $t_\mathrm{ingest}$; by contrast,
$\tau_\mathrm{mb}(t\le t_\mathrm{ingest})>t_\mathrm{ingest}$ for F stars.
The strongest constraints on the parameters of the postulated SHJ population are
provided by the observed $\lambda$ distributions of HJs for the G- and F-star
systems.
In particular, by comparing the number counts of well-aligned HJs in solar-type
stars (corresponding to the first bin in the lower left panel of
Figure~\ref{fig:cool_stars}) with those in the $\lambda>20^\circ$ bins, one
infers $p\sim0.5$: the value of this parameter can be expected to reflect both
the formation rate and radial transport properties of HJs in protoplanetary
disks and the fraction of systems in which disk truncation is either short-lived
or inefficient \citep[e.g.,][]{HerbstMundt05}.
The corresponding data for F stars (lower left panel of
Figure~\ref{fig:hot_stars}) in turn imply an upper bound ($\sim1\,M_\mathrm{J}$)
on $M_\mathrm{SHJ}$:
If the SHJ mass were measurably larger, there would be significantly more
aligned HJs detected in these systems.
Interestingly, this upper bound is consistent with the limit obtained in the SHJ
formation scenario studied by \cite{Rice+08}.
We emphasize that the qualitative results of our model are not sensitive to the
exact values of these (or any of the other) parameters.
Additional data would, however, be useful for better constraining these values.

The host stars also interact tidally with HJs that are still orbiting (and thus
remain observable), which typically have $P_\mathrm{orb}$ values of a few days.
While the impact of this interaction is not as pronounced as that of ingestion,
it can still affect the observed properties of these systems.
This is illustrated in the middle and right bottom panels of
Figures~\ref{fig:cool_stars} and~\ref{fig:hot_stars}, which display,
respectively, the dependence of $\lambda$ and of the projected stellar angular
velocity on the HJ mass for a random selection of $100$~simulated systems.
It was already recognized before that the tidal interaction scenario is
consistent with the apparent dearth of massive planets with either retrograde
orbits or high-$P_*$ hosts:\footnote{
The decrease of $P_*$ with increasing $M_\mathrm{p}$ was noted in the case of
cool stars by \cite{Dawson14}; however, as the data shown in
Figure~\ref{fig:hot_stars} suggest, it also characterizes hot stars.}
This follows from the inverse dependence of $\tau_\mathrm{d}$
(Equation~(\ref{eq:tau})) on $M_\mathrm{p}$, which implies that more massive HJs
should be more efficient at realigning and spinning up their hosts.
The displayed results demonstrate that our model also broadly reproduces these
trends quantitatively, for both G and F stars.

\section{Discussion}

Our interpretation of the good alignment inferred for long-period planets around
cool stars \citep{Mazeh+15} relies on the orbital planes of the ingested SHJ and
of any remaining distant planet having roughly the same orientation, which is
consistent with the picture of giant planets forming in a nearly planar disk and
migrating to the vicinity of the host star.
However, this scenario is not consistent with interpretations of primordial
misalignment in terms of either planet--planet scattering
\citep[e.g.,][]{BeaugeNesvorny12} or a Kozai-Lidov--type interaction with a
misaligned companion \citep[e.g.,][]{FabryckyTremaine07}, as these mechanisms
typically result in different orientations for the orbit of an HJ and those of
more distant planets.
There have already been several proposals in the literature for forming
protoplanetary disks that are misaligned with the protostellar spin axis,
including a warping torque exerted by the stellar magnetic field \citep{Lai+11};
a gravitational torque exerted by a misaligned companion \citep{Batygin12},
possibly amplified by a resonance between the disk-torquing frequency and the
(disk-driven) stellar precession frequency \citep{BatyginAdams13}; a combination
of the above two torques \citep{Lai14,SpaldingBatygin14}; and accretion from a
turbulent interstellar medium \citep{Bate+10,Fielding+15}.
However, these models have not yet been fully evaluated in light of the observed
$\lambda$ distribution of transiting planets \citep[e.g.,][]{CridaBatygin14} and
other observational constraints \citep[e.g.,][]{Watson+11,Greaves+14}.
In this work we adopted a completely random distribution of obliquities for
simplicity, but more data and improved modeling might be able to constrain
$f(\psi_\mathrm{ini})$ in the context of this scenario.
Future steps toward refining our model could include taking account of stellar
evolution and of the effect of dynamical tides, incorporating an inner boundary
condition that represents the stellar magnetosphere into global models of planet
formation and evolution to better predict the properties of SHJs, and
reconciling the SHJ formation picture with the observed distribution of
lower-mass planets \citep[e.g.,][]{Mandell+07,FoggNelson07}.

The clearest observational prediction of this model is the occurrence of highly
misaligned HJs with $P_\mathrm{orb}\lesssim2\,$days around very young stars, but
finding such objects could be rather challenging
\citep[e.g.,][]{Miller+08,Crockett+12}.
There is at present only one claimed detection of a short-period
($P_\mathrm{orb}\approx0.45\,$days), Jupiter-mass planet around a pre-MS
($\sim3\,$Myr), low-mass star \citep{vanEyken+12}, and it has, in fact, been
inferred to be highly misaligned ($\lambda\sim70^\circ$); however, the true
nature of this object is still being debated (e.g., \citealt{Ciardi+15} and
references therein).

\acknowledgements
We are grateful to Dan Fabrycky, Tsevi Mazeh, Fred Rasio, Seth Teitler,
Francesca Valsecchi, and Joshua Winn for fruitful discussions. 
This  work was supported in part by NASA ATP grant NNX13AH56G.

\end{document}